\documentclass[twocolumn]{aa}
\usepackage{graphicx}
\usepackage{txfonts}
\begin{document}
%
   \title{[\ion{O}{iii}] profile substructure in radio--quiet
   quasars\thanks{Based on observations made with ESO Telescopes at
   the La Silla Observatory under programme ID 73.B--0290.}}

   \author{Christian Leipski
          \and
          Nicola Bennert
          }

   \offprints{C. Leipski}

   \institute{Astronomisches Institut Ruhr--Universit\"at Bochum,
              Universit\"atsstrasse 150, D--44780 Bochum, Germany
              \email{leipski@astro.rub.de, nbennert@astro.rub.de}
             }

   \date{Received July 25, 2005; accepted November 4, 2005}

   \abstract{Interactions between the radio jet and the optical emission of
   the narrow--line region (NLR) are a well known phenomenon in Seyfert
   galaxies. Here, we present the study of possible jet--NLR
   interactions in five radio--quiet PG quasars with double or
   triple radio structure.
   High spatial and spectral resolution observations were carried out in the
   H$\beta$--[\ion{O}{iii}]\,$\lambda$5007 wavelength range.
   In all cases, there is evidence for
   [\ion{O}{iii}] profile substructure (shoulders, subpeaks, blueshifted
   ``broad'' components) with different clarity. To measure the
   velocity, line width, intensity, and 
   location of these [\ion{O}{iii}] components, several Gaussians were fitted.
   Often, the substructures are more pronounced close
   to the radio lobes, suggestive of jet--NLR interactions. 
   Our observations support the
   unification scheme in which radio--quiet quasars are assumed to be
   the luminous cousins of Seyfert galaxies. 
   }
\titlerunning{[\ion{O}{iii}] profiles in radio--quiet quasars}
\authorrunning{Leipski \& Bennert}
   \maketitle
%

\section{Introduction}
The narrow [\ion{O}{iii}]\,$\lambda 5007$ emission typical for
active galactic nuclei (AGNs) originates from the ionized narrow--line
region (NLR) gas surrounding the luminous central engine,
most likely an accreting supermassive black hole (BH).\\\indent
For Seyfert galaxies, significant progress has been made in the past
few years in understanding the physics of the NLR and revealed a
close interaction between the radio jet and the NLR gas. Both
the radio--emitting region and the NLR have comparable sizes and
often show a closely matching morphology.
The radio structure is known in many cases to comprise
linear double or triple sources
(Ulvestad \& Wilson \cite{ulvestad84a},\cite{ulvestad84b}; Schmitt
et\,al. \cite{schmitt01}) and the 
emission--line morphologies often show a rich variety such as 
strands or bow--shock like structures (Falcke et\,al. \cite{falcke98};
Whittle et\,al. \cite{whittle04}).\\\indent
Whittle et\,al. (\cite{whittle88}) observed ten Seyfert galaxies with
aligned multiple radio structures and found clear evidence for
[\ion{O}{iii}] profile substructure, often coinciding
with the position of radio lobes. The presence of two
line emitting constituents in the NLR (``ambient
[\ion{O}{iii}]'' which displays an apparently normal rotational velocity field
as well as kinematically displaced ``[\ion{O}{iii}] components'' which are
intimately associated with individual radio lobes) are interpreted as 
clear signatures of the interaction of the radio jet with the
dense surrounding interstellar 
medium, supporting the findings from imaging campaigns.\\\indent
While quasars are assumed to be the luminous cousins of Seyfert galaxies,
there is a significant lack of comparable dedicated quasar NLR studies. 
Most comparable studies of high--luminosity AGN focus on radio--loud
objects, using pure radio or 
emission--line imaging to test radio--NLR alignments.  
There are only a few detailed spectroscopic studies of the jet
gas--interactions of radio galaxies  
(e.g. Clark et\,al. \cite{clark97}, \cite{clark98}) or compact steep--spectrum 
sources (e.g. Gelderman \& Whittle \cite{gelderman94}, Chatzichristou
et\,al. \cite{chatzichristou99}),  
including quasars.
All of these sources have radio luminosities (and therefore suspected jet
energies) that 
are a few orders of magnitude greater than in Seyferts, possibly
altering the interaction scenario significantly.\\\indent
Bennert et\,al. (\cite{bennert02}) performed the first comprehensive HST 
study of structure and morphology of a well--selected
sample of seven radio--quiet Palomar Green (PG) quasars,
observed in [\ion{O}{iii}]. All seven quasars revealed extended
[\ion{O}{iii}] emission. Additionally, Leipski et
al. (\cite{leipski05}) present high sensitivity
radio--maps of radio--quiet  
quasars (RQQ; including those of Bennert et\,al. \cite{bennert02}) and 
Seyfert galaxies to investigate the jet--NLR interaction by comparing
their morphologies: There is a
striking similarity of the radio and [\ion{O}{iii}] images suggesting
equally pronounced jet--NLR interactions in both types of objects.\\\indent
Here, we explore the interaction scenario of 
radio--jet and NLR gas in RQQ in a pilot study of five objects by means of
long--slit spectroscopy.
\begin{table}
\centering
\begin{minipage}[t]{\columnwidth}
\caption{Observations}             
\label{observation}      
\renewcommand{\footnoterule}{}  
\begin{tabular}{cccccc} 
\hline\hline      
Object &  z\footnote{From the peak velocity of the broad H$\beta$
  line} & integration &
p.a. & seeing & ${\rm pc}/^{\prime\prime}$\\ 
\hline                    
PG\,0157+001   & 0.162 & 3600\,s & 106\degr & $<$0\farcs7 & 2750 \\
PG\,1012+008   & 0.186 & 3600\,s & 142\degr & $<$1\arcsec & 3110 \\ 
PG\,1119+120   & 0.049 & 3600\,s & 53\degr  & $<$1\arcsec & 985  \\
PG\,1149$-$110 & 0.048 & 3600\,s & 78\degr  & $<$0\farcs7 & 965  \\
PG\,1307+085   & 0.154 & 4800\,s & 112\degr & $<$0\farcs7 & 2660 \\
\hline 
\end{tabular}
\end{minipage}
\end{table}
\section{Observation and Reduction}
High spatial resolution long--slit spectra
of five radio--quiet PG quasars (PG\,0157+001, 
PG\,1012+008, PG\,1119+120, PG\,1149$-$110, and PG\,1307+085)
were obtained using EMMI 
attached to the Nasmyth B focus of the ESO--NTT.
Observations were made in the H$\beta$/[\ion{O}{iii}] spectral range through 
the nucleus of each quasar with exposure times of $\sim$60 minutes in
total and a 
typical seeing of $<$1\arcsec (see Table~\ref{observation} for details).
The spatial resolution element is 0\farcs3\,pix$^{-1}$,
the nominal spectral  
resolution 0.4\,\AA\,pix$^{-1}$.
The slit width used corresponds to 0\farcs5 on the sky projecting
to a spectral resolution of $\sim$0.9\,\AA~($\sim$ 50 km\,s$^{-1}$ at
5300\,\AA)  
as measured by the full width at half maximum (FWHM)
of comparison lines and night--sky lines.
The long--slit used corresponds to 200\arcsec~in the sky and 
was orientated along the position angle (p.a.) 
of the maximum extent in the [\ion{O}{iii}] or radio image as
determined from the images of Bennert et.\,al (\cite{bennert02}) and
Leipski et.\,al (\cite{leipski05}).\\\indent
Standard reduction including bias subtraction, flat--field correction,
cosmic--ray removal, wavelength, and flux--calibration was performed
using the ESO--MIDAS software (version Feb. 04). \\\indent
To check the contribution of any \ion{Fe}{ii} emission at
4924\,\AA~and 5018\,\AA, we followed the procedure of Peterson
et\,al. (\cite{peterson81}), using the $\lambda4570$ blend as an
estimate of the \ion{Fe}{ii} emission. We find that the \ion{Fe}{ii}
emission is neglible. Moreover, a significant contribution of
\ion{Fe}{ii} emission at 
4924\AA~and 5018\AA~would make the two [\ion{O}{iii}]  
lines asymmetric in the opposite senses, what is not observed here.\\\indent
Two to three rows were averaged from the frames  
according to the seeing to enhance the S/N without loosing any spatial
information. Thus, each ``resulting row'' 
corresponds to 0\farcs7 or 1\farcs0 along and 0\farcs5 perpendicular
to the slit direction, respectively. 
Along the ``spatial axis'' of the CCD, we identified the
``photometric center'' (that we choose as ``zero'' on the spatial
scale) with that spectral row of the CCD where the
continuum is at maximum
(``central row'').
The linear scales (${\rm pc}/^{\prime\prime}$) were calculated  using
the velocity relative  
to the 3K background, H$_0$ = 71 km\,s$^{-1}$\, Mpc$^{-1}$ 
and a world model in agreement with the recent
results of the Wilkinson Microwave Anisotropy Probe (Bennett
et\,al. \cite{bennett03}).\\\indent
Velocities, line widths, and intensities of the different
[\ion{O}{iii}] components were measured as a function
of distance from the nuclei by fitting Gaussians to the line profile.\\\indent
In a first attempt, we tried to fit the [\ion{O}{iii}]$\,\lambda5007$
line with the least number of gaussians. This set of gaussians was
extended to also fit [\ion{O}{iii}]\,$\lambda4959$ (with a fixed flux
ratio of 3:1).
The parameters of the successful fit were then used to fit H$\beta$,
taking into account only those components of the [\ion{O}{iii}] fit with a
significant contribution to the H$\beta$ line (peak values as free
parameters). Additionally, we included a broad component without
constraints.\\\indent 
\section{Results and Discussion}
\begin{figure}
         \includegraphics[angle=0,width=8.5cm]{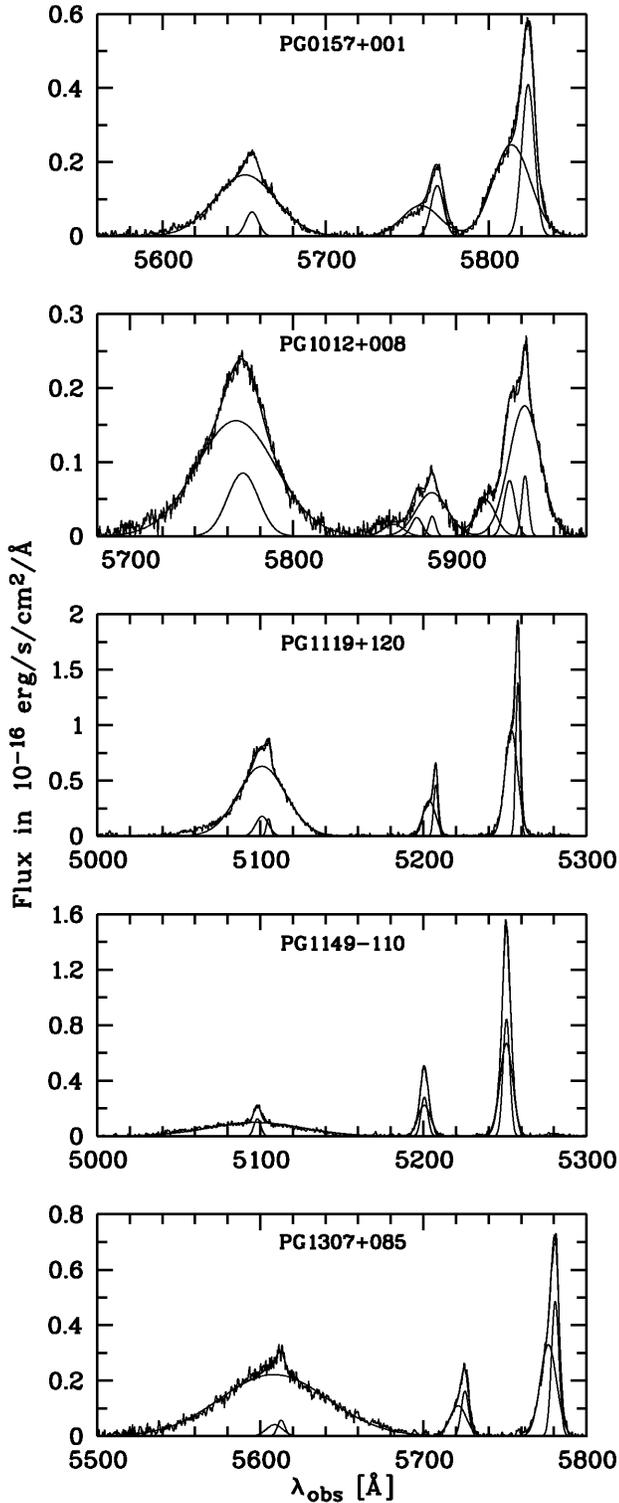}
         \caption[]{\label{all_fits}Central spectra of the five
         PG quasars. The observed spectrum is shown,
         overlayed with the fitted components as well as with the
         total fit.}
\end{figure}

In the following, we present the results on an object--by--object
basis. The central spectra with the fitted components are shown in
Fig.\,\ref{all_fits}. 
We were not able to estimate the excitation (as measured
from the [\ion{O}{iii}]/H$\beta$ ratio) for all
components due to dilution by the broad
H$\beta$ line.
While the broad emission is especially strong in the nuclear spectrum, we 
were limited by a lack of signal in the outer spectra.
Moreover, in all five quasars the 
[\ion{O}{iii}] line is (significantly) stronger than H$\beta$
(Fig.\,\ref{all_fits}) and we were often limited by a noisy (or even
absent) H$\beta$ line. For those [\ion{O}{iii}] components which could be
identified and fitted in the H$\beta$ line, we give the ratios in the text.
The different components have on average high [\ion{O}{iii}]/H$\beta$
values ($\sim$\,5--8), consistent with photoionization by the AGN.
\subsection{PG\,0157+001}
\begin{figure}
         \includegraphics[angle=0,width=8.5cm]{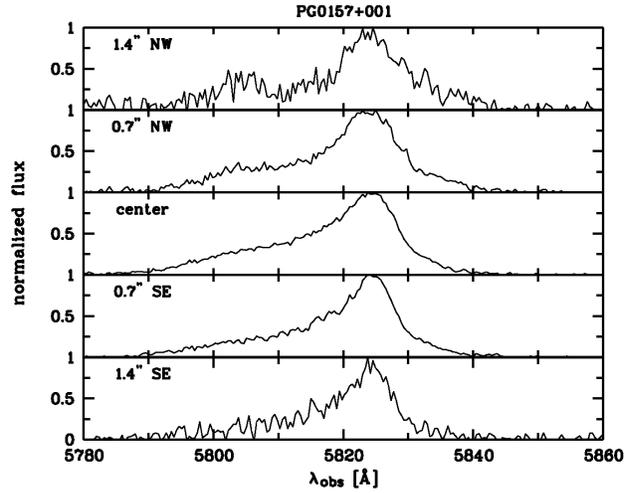}
         \caption[]{\label{pg157} Spatial sequence of spectra of \object{PG0157+001}.}
\end{figure}

This radio--quiet quasar at a redshift of z\,=\,0.163
(Tab.\,\ref{observation}) comprises a triple radio source
and shows exceptional structures in the HST [OIII] emission--line
images (Bennert et\,al. \cite{bennert02}). Being
well aligned with the radio structure, the [OIII] emission 
displays a prominent bow--shock like arc on the western side of the
nucleus suggesting
strong jet--NLR interaction. On the eastern side, an emission blob is
lying at the same distance from the nucleus as the radio knot
($\sim$0\farcs7) but slightly displaced to the north ($\sim$0\farcs5)
(Leipski et\,al. \cite{leipski05}).
On larger scales ($\sim$6\arcsec), Stockton \& MacKenty (\cite{stockton87})
show extended emission--line features in their ground--based
observations that closely resembles the structures seen on small
scales.\\\indent 
We aligned the long--slit with these larger
structures and indeed detect extended emission on similar scales. However,
the low S/N does not allow a detailed line fitting.
Nevertheless, some basic parameters can be determined: The excitation
of the north--western emission is twice as high 
as that of the south--eastern region ([\ion{O}{iii}]/H$\beta=7.2$
vs. 3.3). There is a small velocity shift between the
south--eastern and the north--western extended emission
($\sim$$80\,$km$\,$s$^{-1}$).
It is not clear whether this velocity difference can be interpreted as
galactic rotation or is due to outflowing gas with the western region
pointing towards us.\\\indent 
In the central part (inner $\sim$3\arcsec), we study
radially varying properties. 
The [\ion{O}{iii}] line can be best fitted by a two--component
system (Fig.\,\ref{all_fits}). As the blue--shifted component is not
seen in the H$\beta$ 
line, it indicates a high excitation
([\ion{O}{iii}]/H$\beta$\,$\geq$\,11) of this very broad 
($\sim$$1400\,$km$\,$s$^{-1}$) feature. As can be seen in 
Fig.\,\ref{pg157}, this broad [\ion{O}{iii}] component is most
prominent in the center, slowly fades out to the south--east, and evolves into
a pronounced and much narrower ($\sim$$560\,$km$\,$s$^{-1}$) secondary peak
$\sim$1\farcs4 from the center in the north--west at a blueshifted
velocity of $\sim$$1200\,$km$\,$s$^{-1}$. This 
position corresponds to the bow--shock like structure in the
emission--line gas distribution (Bennert et\,al. \cite{bennert02}),
also coinciding with the radio knot observed by Leipski
et\,al. (\cite{leipski05}). These findings strongly suggest an
interaction of the radio ejecta 
with the emission--line gas, which is swept--up and driven outwards
(towards the observer) by the jet, thus creating the bow--shock.\\\indent
Note that additional components may possibly be seen in
the [\ion{O}{iii}] emission--line profile of \object{PG0157+001}
(e.g. a red wing or a 
component in between the two fitted ones visible at 1\farcs4 north--west,
Fig.\,\ref{pg157}). These components were included in the fit on an
individual row--by--row basis, but no consistent fitting scenario could be
found for all rows. This indicates that the radial
variations of the line profile are more complex.
\subsection{PG\,1012+008}
The quasar \object{PG1012+008} is merging with a nearby
companion (Bahcall et\,al. \cite{bahcall97}). This complicates the
interpretation of the profile
substructures which may be either caused by the merger or by
the radio jet. 
The radio structure of \object{PG1012+008} consists of a central
source with a $\sim$1\farcs5 long bent jet to the south--east. The
emission--line gas is distributed rather uniformly, but the shape
of the isophotes resembles that of the radio structure (Leipski
et\,al. \cite{leipski05}).\\\indent
In our spectra, the object exhibits a
wealth of structure in the [\ion{O}{iii}] profile
(Fig.\,\ref{all_fits}). A blue
component is displaced by nearly $1000\,$km$\,$s$^{-1}$ with respect
to the peak suggesting  
a fast outflow. While this component is detected in the inner 3\arcsec,
it is most prominent in the center compared to the peak of the total
profile. The observed emission is
confined to these 3\arcsec~and the profiles do not vary significantly
with distance from the center. However, the
blue asymmetry seen in the central [\ion{O}{iii}] profile
becomes a red asymmetry towards the south of the nucleus, indicative
of material flowing in south--eastern direction. This coincides with
the direction of the radio jet. Moreover, the [\ion{O}{iii}]
emission--line image of this source shows elongated 
isophotes in the very same direction. Whether the observed outflow can be
attributed to a radio jet originating in the active nucleus or is
influenced by the merger event cannot be answered at present. We
suggest that the outflow itself is powered by the active nucleus but
the extended distribution of the
gas seen in the [\ion{O}{iii}] image as well as the "bent" shape of
the jet is due to the interaction of the two galaxies.
\subsection{PG\,1119+120}
\object{PG1119+120} shows radio structures of $\sim$5\arcsec\,diameter.
On the eastern side of the nucleus, a jet
sharply bends to the north at a distance of 2\farcs5, while a single radio knot
is detected on the western side (Leipski et\,al. \cite{leipski05}).
We extract spectra on similar scales. The central
spectrum of this source reveals a two--component 
[\ion{O}{iii}] line with both components visible in the
H$\beta$ line as well. 
The radial variations are most obvious towards the
south--west: While the narrower, red component has a stable
[\ion{O}{iii}]/H$\beta$ ratio ($\sim$\,8), the of the blue--shifted 
broader component increases by 40\% from the nucleus out to $\sim$1\farcs5 (SW)
([\ion{O}{iii}]/H$\beta$\,$\sim$\,5.3 to 8.7). Similarly, while the flux
in both components 
equals at that distance (which coincides with the position of the
western radio knot), the red 
component is stronger in the nucleus by a factor of 2. The presence of
the radio knot and the increasing flux of the blue--shifted component
indicate strong jet--NLR interaction in the south--west of the
nucleus.\\\indent 
According to its velocities, the narrower, red component seems to represent
ambient [\ion{O}{iii}] emission that follows galactic rotation, while
the (variable) blueshifted [\ion{O}{iii}] component is most likely 
attributed to gas motions caused by the radio ejecta. Interestingly, no
remarkable [\ion{O}{iii}] substructures are visible at the location of
the north--eastern radio jet that is much more pronounced in the radio
regime.    
\subsection{PG\,1149$-$110}
Although this object shows extended structures in the radio regime, no
significant [\ion{O}{iii}] profile substructure can be seen
(Fig.\,\ref{all_fits}). Nevertheless, two Gaussians are needed to fit
the observed profile due to broader wings, a phenomenon already known
from Seyfert galaxies (e.g. Whittle et\,al. \cite{whittle85}, Ver\'on--Cetty
et\,al. \cite{veron01}, Schulz \& Henkel \cite{schulz03}, Bennert
et\,al. \cite{bennert04}).
We were not able to extract more than 3 spectra
(diameter\,$\sim$2\arcsec) at a S/N $>$ 3.
\object{PG1149$-$110} shows that the
presence of extended radio emission alone does not necessarily imply
significant 
profile substructure in the [\ion{O}{iii}] lines. This may indicate that
the extended radio emission suggestive of radio jets
expands nearly perpendicular with respect to the observer resulting in
an absence of pronounced [\ion{O}{iii}] components. However, there are signs
of a red asymmetry at $\sim$\,1\arcsec~east, at the
location of the eastern radio knot (Leipski et\,al. \cite{leipski05})
indicating interaction.
\subsection{PG\,1307+085}
\begin{figure}
         \includegraphics[angle=0,width=8.5cm]{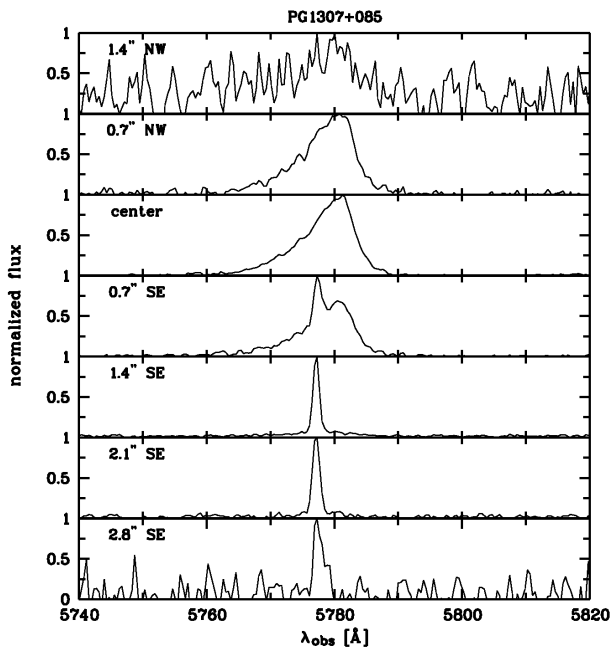}
         \caption[]{\label{pg1307} Spatial sequence of spectra of \object{PG1307+085}.}
\end{figure}

The host galaxy of \object{PG1307+085} is classified as a small
early--type galaxy (Bahcall et\,al. \cite{bahcall97}) and, like
\object{PG0157+001}, this is a quasar in
our sample that shows remarkable spatially varying [\ion{O}{iii}]
profile substructure: In the center and towards the north--west, the
[\ion{O}{iii}] profile consists of a narrow central component with a
broad blue shoulder (Fig.\,\ref{all_fits}). From 0\farcs7 to 2\farcs8
south--east, an additional narrow component appears which is not visible in
the other rows (Fig.\,\ref{pg1307}). The presence of this component in
the central rows is difficult to assess because of the presence of the
broad component which is much stronger.
However, on the other side of the nucleus at 0\farcs7
north--west, it may reflect in a 
weak shoulder bluewards of the peak of the line profile. We identify this
component with ambient [\ion{O}{iii}] gas in the host galaxy. The
emission--line of the ambient gas is very narrow
($\sim$\,1\,\AA\,$\sim\,$$69\,$km$\,$s$^{-1}$) and shows no
significant velocity shift. 
\section{Conclusions}
In all five observed radio--quiet PG quasars, we find evidence for
[\ion{O}{iii}] 
profile substructure with different clarity. Often, this substructure
is pronounced at the location of the radio lobes. This suggests that
at least part of the emission--line gas is intimately associated with
individual radio lobes, indicating a close NLR--jet
interaction. However, the presence of radio lobes does not
necessarily lead to the 
presence of pronounced [\ion{O}{iii}] profile substructure. The
orientation of the radio--jet with respect to the observer seems to be a
crucial factor for enhanced profile substructure.
Our results are comparable with similar
studies of Seyfert galaxies (e.g. Whittle et.\,al \cite{whittle88})
and support the  unification scheme in which
radio--quiet quasars are assumed to be the luminous cousins of
Seyferts. However, the
excitation as measured from the [\ion{O}{iii}]/H$\beta$ ratio turns
out to be generally lower for our quasars ($\sim$\,5--8) than for
Seyfert galaxies ($\sim$\,10--12, Whittle et\,al. \cite{whittle88}). 
\begin{acknowledgements}
C.L. was supported by Sonderforschungsbereich SFB\,591
  ``Universelles Verhalten gleichgewichtsferner Plasmen'' der
  Deutschen Forschungsgemeinschaft. N.B. is grateful for financial
support by the ``Studienstiftung des deutschen Volkes''. We also
  acknowledge the helpful comments of the anonymous referee.
\end{acknowledgements}

\end{document}